\documentclass[prb,aps,preprint,showpacs]{revtex4-1}
\usepackage{graphics}
\usepackage{epsfig}
\usepackage{graphicx}
\usepackage{float}

\begin{document}

\title{Atomic scale chemical fluctuation in LaSrVMoO$_6$: A proposed halfmetallic antiferromagnet}

\author{Somnath Jana,$^{1}$ %
Vijay Singh,$^{2}$ %
S. D. Kaushik,$^{3}$ %
Carlo Meneghini,$^{4}$ %
Prabir Pal,$^{5}$ %
Ronny Knut,$^{5}$ %
Olof Karis,$^{5}$ %
Indra Dasgupta,$^{1,2}$ %
Vasudeva Siruguri,$^{3}$ %
Sugata Ray$^{1,6,\star}$}

\affiliation{$^1$Centre for Advanced Materials, Indian Association for the Cultivation of Science, Jadavpur, Kolkata 700 032, India\\ %
$^2$Solid State Physics, Indian Association for the Cultivation of Science, Jadavpur, Kolkata 700 032, India\\ %
$^3$UGC-DAE-Consortium for Scientific Research Mumbai Centre, R5 Shed, Bhabha Atomic Research Centre, Mumbai 400085, India\\ %
$^4$Dipartimento di Fisica Universit\'a di ``Roma Tre" Via della vasca navale, 84 I-00146 Roma, Italy\\ %
$^5$Uppsala Univ, Dept Phys \& Mat Sci, S-75121 Uppsala, Sweden\\ %
$^6$Department of Materials Science, Indian Association for the Cultivation of Science, Jadavpur, Kolkata 700 032, India}

\begin{abstract}
Half metallic antiferromagnets (HMAFM) have been proposed theoretically long ago but have not been realized experimentally yet. Recently, a double perovskite compound, LaSrVMoO$_6$, has been claimed to be an almost real HMAFM system. Here, we report detailed experimental and theoretical studies on this compound. Our results reveal that the compound is neither a half metal nor an ordered antiferromagnet. Most importantly, an unusual chemical fluctuation is observed locally, which finally accounts for all the electronic and magnetic properties of this compound.
\end{abstract}

\maketitle


Half metallic antiferromagnet (HMAFM), proposed by van Leuken and de Groot,~\cite{Leuken} have attracted enormous attention in recent times as it offer an unique possibility of realization of an exotic state that has large degree of spin polarization of conduction electrons, but vanishing macroscopic magnetic moment. Unlike a conventional antiferromagnet, in a HMAFM the nullification of magnetic moment occurs between the anti-parallel spins of same magnitude, residing on different ions of distinct symmetry, and as a result a perceptible exchange splitting between the up- and down-spin channels is observed. Due to this unusual situation, HMAFM materials are also sometime termed as `compensated half-metals' (CHM).~\cite{pickett_PRB} A large number of systems, {\it e.g.} metal alloys,~\cite{heusler_PB,alloy_JAP, alloy_JMMM} metal pnictides,~\cite{pnictide_JPCM} or even organic polymers,~\cite{organic_JCP} have been theoretically predicted to be possible HMAFM candidates, but none of them could be experimentally realized so far. In this respect, double perovskite materials ($AA^\prime$$BB^\prime$O$_6$) have turned out to be one of the popular searching grounds,~\cite{pickett,park,guo} because of the possibility of tuning the $B,B^{\prime}$ cation occupancy to achieve desired properties. But, even this search was not successful.~\cite{Androulakis,date_PRB} Therefore, existence of a true HMAFM appeared really elusive until Uehara {\it et al.}~\cite{Uhera} proposed the curious case of LaSrVMoO$_6$ (LSVMO). LSVMO has been suggested to be a HMAFM system by considering antiparallel alignment of alternately occupied V$^{3+}$ (3$d^2$) and Mo$^{4+}$ (4$d^2$) spins at the $B$-site of the perovskite structure. Although a recent calculation~\cite{park2} predicted LSVMO to be a half-metallic {\it ferrimagnet}, based on the magnetic and transport measurements, it was claimed that the system can indeed become HMAFM for a perfectly $B$-site ordered state.~\cite{Uhera} Recently, this suggestion has been endorsed very strongly in an another report,~\cite{Gotoh} where simultaneous presence of a long range AFM order and as large as 50\% spin polarization in LSVMO has been claimed. However further careful experimental studies are warranted on this material, in view of the fact that the $B$-site disorder and the true local structure may play an important role here, as has been proved in case of other double perovskites.~\cite{Meneghini}

We have carried out detailed experimental studies on polycrystalline LaSrVMoO$_6$, which along with electronic structure calculations revealed that the ground state of this compound is far from being half-metallic and also there is no long-range AFM order. Instead, x-ray absorption fine structure (XAFS) experiments indicated presence of unusual chemical fluctuations within small spatial regions of the material, which finally succeeded to explain the deceptive HMAFM behavior of this system.

Polycrystalline LSVMO was synthesized by the solid state route following earlier reports.~\cite{Gotoh} The phase purity of the sample was checked by XRD using a Bruker AXS: D8 Advance x-ray diffractometer, while the magnetic measurements were carried out in a Cryogenics Physical Property Measurement System and in a Quantum Design SQUID magnetometer. Low energy x-ray absorption (XAS) and x-ray magnetic circular dichroism (XMCD) measurements were carried out at I1011 beamline of the Swedish synchrotron facility MAX-lab, Lund. All the XAS and XMCD spectra were measured by recording the total electron yield. Every time the sample was heated to 250 $^{\circ}$C {\it in-situ} to remove physisorbed surface species. As LSVMO is quite stable towards additional oxidation beyond the outmost surface layer and the the probing depth is within 50-100 \AA, we could probe the bulk properties. Neutron powder diffraction (NPD) measurements were carried out at Focusing Crystal Diffractometer set up by the UGC-DAE Consortium for Scientific Research Mumbai Centre at the Dhruva reactor, Mumbai (India).~\cite{vasu1} NPD data were analyzed using Rietveld technique and was refined using the FullProf program.~\cite{vasu2} Mo $K$-edge XAFS measurements were performed at the BM08-GILDA beamline at the ESRF (Grenoble). Electronic structure calculations were performed in a plane-wave basis set using the projector augmented wave (PAW) ~\cite{paw} method as implemented in Vienna ab-initio Simulation Package (VASP).~\cite{vasp1, vasp2} An energy cut off of 500 eV for the plane wave expansion of the PAW's was employed in our calculations. The exchange correlation part was approximated in generalised gradient approximation (GGA) including a Hubbard onsite $d-d$ Coulomb interaction $U$ = 4~eV and onsite exchange interaction $J$=0.9~eV for V.

In Fig. 1(a), we show the room temperature NPD data from LSVMO, along with the refined pattern (line through data points) and the difference spectrum (lower line). Interestingly, instead of the previously assumed cubic structure,~\cite{Uhera, Gotoh} an orthorhombic $Pnma$ structure is found to be the correct one (Table I). It is to be noted that grossly our powder XRD data matches well with earlier reports (inset to Fig. 1(a)), while a critical look at the NPD pattern reveals the discrepancy. For example, the reflections (122) and (221), marked with asterisks in Fig. 1(a), are not allowed in the $Fm3m$ space group and are very weakly present in the XRD pattern due to the low structure factors of these reflections for x-rays. Another important point to note is that the refinement result shows nearly complete $B, B^{\prime}$-site disorder in this material, which could not be improved by repeated thermal treatments. Resistance ($R$) {\it vs.} temperature ($T$) data is shown in Fig. 1(b) with 0 and 5 Tesla magnetic fields down to 2 K.
\begin{center}
\begin{table} [h]
\caption{Structural parameters\protect\footnotemark[1] after Rietveld refinement of NPD spectrum of LaSrVMoO$_6$ at room temperature.}
\begin{tabular}{|c|c|c|c|c|c|}
\hline
Atom & $x$ ({\AA}) & $y$ ({\AA}) & $z$ ({\AA}) & $B_{iso}$ & Occ.\\
\hline
La/Sr & 0.01498 & 0.25 & 1.00057 & 0.75 & 0.5\\
\hline
Mo/V & 0.5 & 0 & 0 & 0.5 & 0.5\\
\hline
O1 & 0.49008 & 0.25 & 0.06088 & 0.96 & 0.5\\
\hline
O2 & 0.26357 & 0.02248 & 0.73265 & 0.9107 & 1.0\\
\hline
\end{tabular}
\footnotetext[1]{Space group: $Pnma$, $a$~=~5.5779~{\AA}, $b$~=~7.8888~{\AA}, $c$~=~246.075~{\AA}, V~=~5.59225~{\AA}$^3$, $\alpha$~=~$\beta$~=~$\gamma$~=~90$^{\circ}$, $\chi^2$~=~3.17, $R_p$~=~2.95\%, $R_{wp}$~=~3.80\%, $R_{exp}$~=~2.13\%.}
\end{table}
\end{center}
These $R$($T$) data clearly justifies the metallic nature of the sample. The temperature evaluation of the magnetization ($M(T)$ in Fig. 1(c)) shows a weak peak around 117 K, followed by a monotonic increase at the low temperature region both in zero field cooled (ZFC) and field cooled (FC) measurements, evidencing a paramagnetic behavior at low temperatures. The divergence in ZFC and FC curves already below 300~K indicates magnetic metastability in the system, which is very different from a simple antiferromagnet, as proposed before.~\cite{Uhera, Gotoh}  Inset to Fig. 1(c) shows $M$ {\it vs} $H$ (field) at 5 K which is almost linear.
\begin{figure}
\begin{center}
\resizebox{8cm}{!}
{\includegraphics*[28pt,90pt][568pt,736pt]{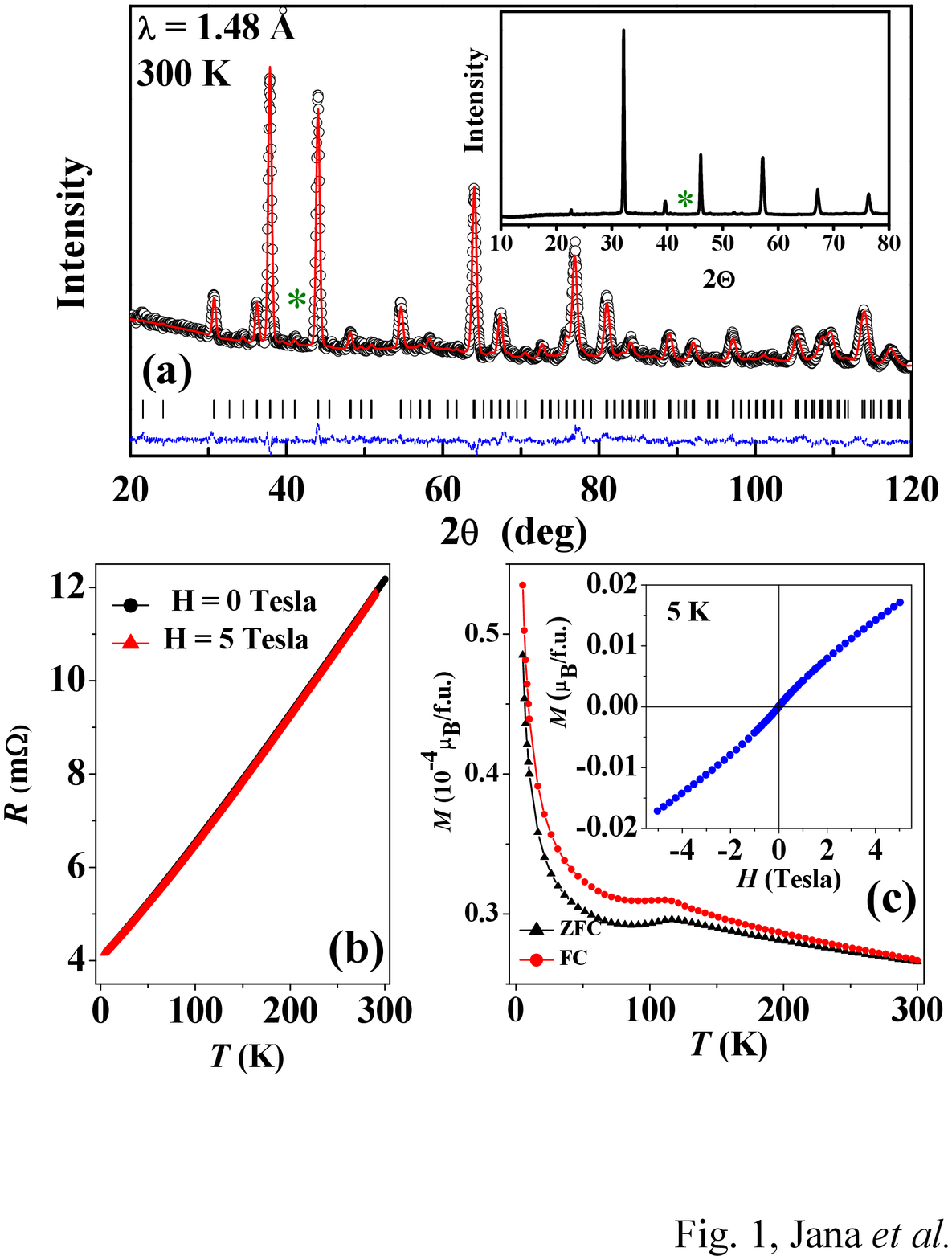}}
\caption{(color online) (a) Neutron powder diffraction from LSVMO at 300 K showing the observed
data (open circles), calculated pattern (solid line over the data points), difference pattern (dashed line) and Bragg positions (vertical black lines). The XRD pattern has been shown as an inset. The $R$($T$) data is shown in (b), while the $M$($T$) data is presented in (c). The inset to (c) shows the $M$($H$) curve at 5~K.}
\end{center}
\end{figure}
Although our results presented so far completely reproduce the results in previous reports,~\cite{Uhera, Gotoh} they do not establish a HMAFM state of the system. To understand the electronic and magnetic structures of the compound better, XAS and XMCD measurements have been performed on LSVMO. XAS of V $L$ and O $K$-edges collected at room temperature are shown in Fig. 2(a) along with the reference spectra from V$_2$O$_3$, VO$_2$ and V$_2$O$_5$, adopted from earlier reports.~\cite{Sara Nordlinder} A close look at the finer structures (indicated by arrows) clearly establishes that vanadium is in $3+$ oxidation state. The Mo $M$-edge XAS (not shown) spectrum is featureless and does not provide useful information regarding the Mo charge state.
\begin{figure}
\begin{center}
\resizebox{8cm}{!}
{\includegraphics*[32pt,158pt][565pt,740pt]{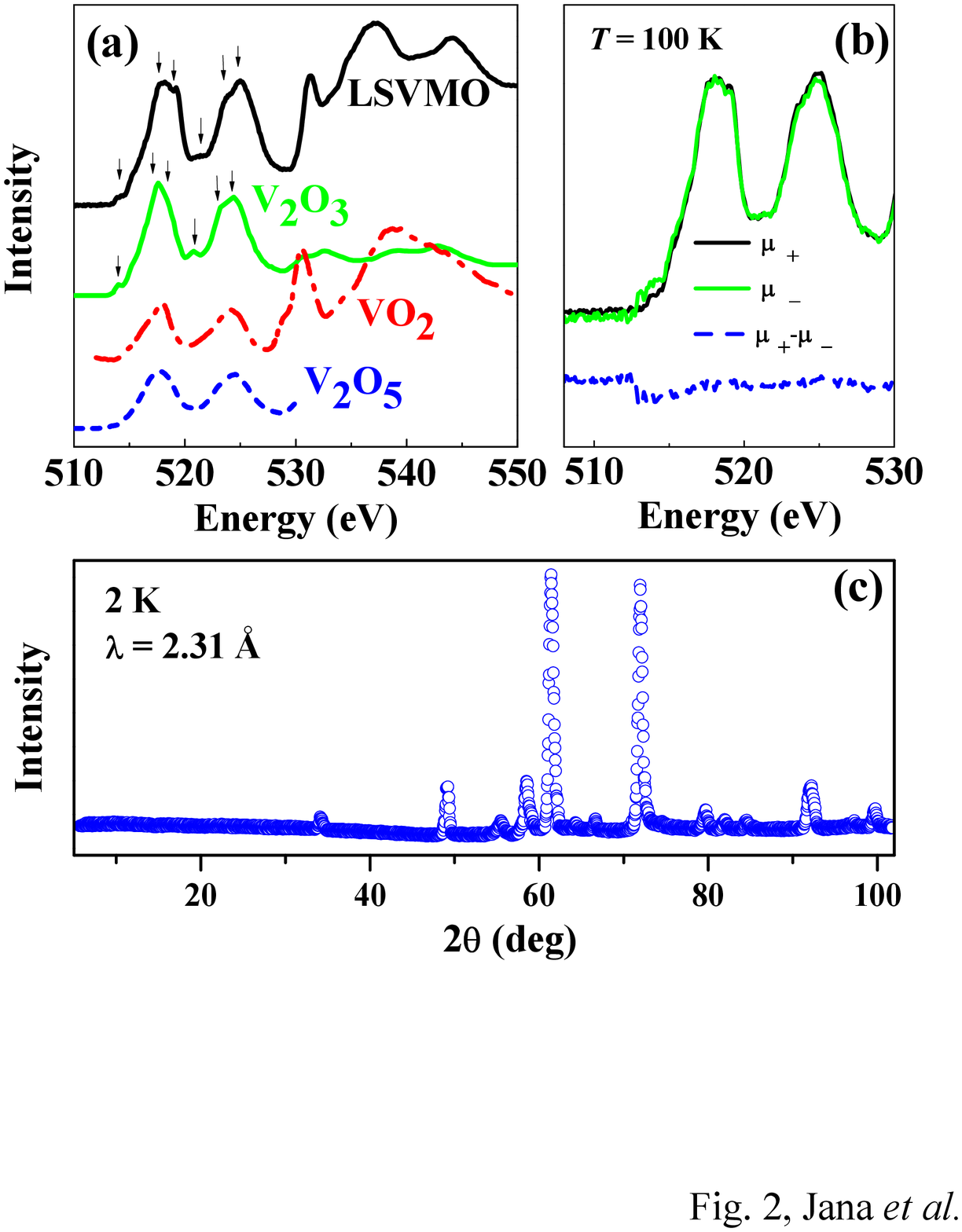}}
\caption{(color online)The V $L_{2,3}$ and O $K$-edge XAS spectra from LSVMO and few other V $L_{2,3}$-edge spectra from standard literature are shown in panel (a). (b) The $\mu_-$, $\mu_+$, and difference spectra from LSVMO. The neutron diffraction pattern at 2 K is shown in panel (c).}
\end{center}
\end{figure}
Now, for a true HMAFM state in LSVMO, compensation of spin moments could only be realized by considering AFM coupling between ferromagnetic V$^{3+}$ and ferromagnetic Mo$^{4+}$ sublattices, and therefore, clear XMCD signal from the polarized V $L$-edge spectra can be expected. But surprisingly, no XMCD signal at vanadium $L$ edge with reverse polarizations at 100~K with a field of 0.3 tesla was observed (see Fig. 2(b)). This observation is in accord with the disordered structure and clearly refutes the HMAFM model, proposed for a perfectly ordered structure. However, even though it is observed that LSVMO is not a half metal, a long-range AFM order below 117 K is still conceivable. Therefore, to check the reliability of this proposed AFM transition, we performed NPD experiments on the sample also at 2 K (Fig. 2(c)). However, it becomes immediately clear from the spectrum that there is no long range magnetic order in the system even at 2 K, as neither any superlattice reflection nor any enhanced intensity at the crystalline Bragg peaks are found throughout the angular range, especially at the lower Bragg angles. Therefore, contrary to previous claims, LSVMO neither possesses the proposed half metallic structure nor has any long range antiferromagnetism, and its observed properties remain enigmatic.

In order to achieve insight about the local chemical structure, which might be critical, XAFS measurements were performed on LSVMO. Fig. 3(a) and 3(b) show the Mo $K$-edge XAFS data, and its Fourier transform along with the respective best fit spectra. A good quality refinement was achieved considering reduced number of Mo neighbor shells up to about 4~\AA. The structural parameters obtained from XAFS analysis are presented in Table II. In case of perfect ordering at $B$, $B^\prime$ sites, every Mo ion has six V ions as the nearest neighbors (Mo-O-V), while for a completely disordered structure, statistically the number of Mo-O-V connectivity should come down to only 3 and consequently, 3 Mo-O-Mo bonds are expected. However, in an average only 2.4 V ions and 3.6 Mo ions are found as the nearest neighbor of a central Mo ion. Moreover, this central Mo ion only has 2.2 La and as many as 5.8 Sr neighbors, instead of 4 La and 4 Sr, as expected for a homogeneous structure.
\begin{figure}
\begin{center}
\resizebox{8cm}{!}
{\includegraphics*[0pt,0pt][236pt,168pt]{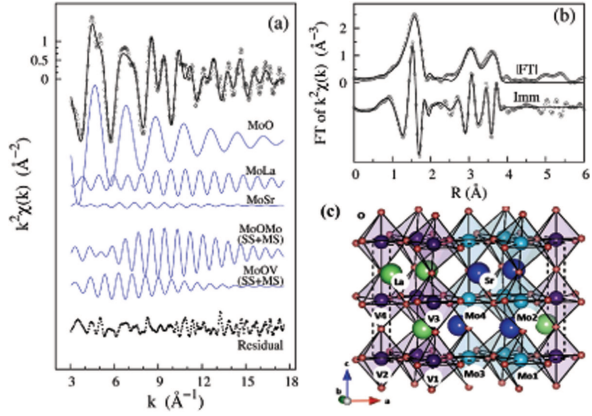}}
\caption{(color online) (a) The k$^2$ weighted XAFS data (open circles) and best fit (black lines) are presented. The partial contributions (blue online) and the residual (k$^2_{exp}$ - k$^2_{th}$) (black dots) are shown, vertically shifted for clarity. (b) Fourier transform of experimental (open circles) and theoretical (dark lines) curves; the moduli ($\mid$FT$\mid$) and imaginary parts (Imm) are vertically shifted for the sake of clarity. In (c) a possible structure based on XAFS results is shown.}
\end{center}
\end{figure}
\begin{table}[h]
\caption{Mo-$K$ edge XAFS results. The coordination numbers are constrained to crystallographic structure. The absolute misfit between experimental data and best fit is $R^2=0.077$.}
\begin{tabular}{|l|c|c|c|c|}
\hline
Shell   & N & R(\AA) & $\sigma^2 (\times 10^3$\AA$^2$) & For calc. \\
\hline
O      & 6       & 1.998(5)  & 2.3(3) & 6\\
MoLa   & 2.2     & 3.47(2)   & 3.4(5) & 2\\
MoSr   & 5.8(4)  & 3.46(2)   & 4.3(5) & 6\\
MoOMo  & 3.6(3)  & 3.97(2)   & 4.8(6) & 4\\
MoOV   & 2.4     & 3.98(2)   & 6.9(6) & 2\\
\hline
\end{tabular}
\end{table}
Therefore, these results reveal an unusual segregation of every Mo(V) ion towards another Mo(V) ion as the nearest neighbor, and towards Sr(La) ion as its next nearest neighbor. In such a scenario, one assumption could be a large scale phase separation of LSVMO into LaVO$_3$ and SrMoO$_3$ type phases but such a possibility is not supported from the diffraction studies. Also, the absence of any AFM order, confirmed by low temperature NPD data, rules out the formation of large scale LaVO$_3$ phase. Thus, only a picture of an intrinsic chemical fluctuation within very small spatial domains, giving rise to very local LaVO$_3$ and SrMoO$_3$-like regions (Fig. 3(c)), could be evoked to explain the observed XAFS results, while all the physical properties must originate from such a structure. The vanadium spins will mostly align antiferromagnetically as in LaVO$_3$, which is an AFM insulator with $T_N$ = 140 K.~\cite{LaVO3} But, any long range magnetic order will be lost because vanadium rich regions are separated by SrMoO$_3$ like regions which is a paramagnetic metal with extremely low electrical resistivity.~\cite{SrMoO3-mag, SrMoO3} Therefore, overall the system can preserve the metallic nature as conduction can occur through a percolation path made from SrMoO$_3$ like regions, while the low temperature rise in the susceptibility (Fig. 1(c)) might be reminiscent of SrMoO$_3$-like paramagnetism. Furthermore, we have performed detailed {\it ab-initio} electronic structure calculations to verify these speculations.

In contrast to an earlier report,~\cite{park2} all the electronic structure calculations presented here are carried out in the orthorhombic structure with experimental lattice parameters, as obtained from the NPD studies. We shall first discuss the results for the ordered structure with two kinds of magnetic ordering, (a) G-type ordering where neighboring V and Mo are antiferromagnetically coupled, (b) A-type ordering where V and Mo ions in the plane are ferromagnetically ordered but neighboring planes are antiferromagnetically coupled. Our calculations reveal that G-type state is most stable ($E_{A}$-$E_{G}$ = 70 meV/f.u.) similar to ref. 14. While A-type magnetic configuration has vanishing total magnetic moment, the same for the G-type is 0.85 $\mu_{B}$, with 1.87 at the V site and -0.87 at the Mo site. This suggests that the V ions possess a valence of $3+$ ($d^{2}$) while the small exchange splitting makes both the spin states occupied for the Mo ions, giving rise to a lower moment. Next, we considered a super cell with four formula units keeping the local coordination of V and Mo very similar to that suggested by XAFS measurements (see Table II). Accommodation of the local co-ordination of Mo and V in the supercell (SC) imposes severe constraint on the number of local structures that are possible for the supercell employed and therefore the present SC calculation forms a typical model for the description of correlated disorder present in LSVMO.

In order to check whether the V spins prefer to align antiparallely with each other, we considered two kinds of spin arrangements for a site occupied by V (Mo), (A) all its nearest neighbors have opposite spin alignment and (B) all its nearest V (Mo) neighbors have parallel spin alignment, while the nearest Mo(V) neighbors have opposite spin alignment. From the results (Table III), we gather that the configuration (A) has the lowest energy, confirming anti-parallel alignment of the neighboring V spins. This is consistent with the absence of XMCD signal. In the disordered case, the energy difference between these two structures becomes even less, suggesting long range order is hardly favorable for the real system. The magnitudes of the magnetic moments at the V and Mo sites, presented in Table III, clearly reveal that LSVMO is not a HFAFM. The magnetic moment at the V site is consistent with its valence of $3+$ and the small (0.3 $\mu_{B}$) moment at Mo site may be attributed to its  very weak exchange splitting.
\begin{table}[h]
\caption{Calculated total energy and magnetic moments for LSVMO in magnetic structure (A).}
\begin{tabular}{|c|c|c|c|c|c|c|}
\hline
Conf & \multicolumn{4}{|c|}{Local Mom. ($\mu_B$)}& Tot Mom & $\Delta$E \\
 & \multicolumn{4}{|c|}{}& ($\mu_B$) & (meV/f.u)\\
\hline
  & V1,4 & V2,3 & Mo1,4 & Mo2,3 & &  \\
\hline
$\rm {{\bf{A}}}$   & 1.84 & -1.84 & 0.30  & -0.30  & 0.00 & 0  \\
\hline
$\rm {{\bf B}}$ & 1.87 & 1.87 & -0.26 & -0.26 & 1.57 & 22 \\
\hline
\end{tabular}
\end{table}

In summary, it is experimentally shown that the claimed HMAFM LaSrVMoO$_6$ is neither a half metal nor a long-range antiferromagnet. Interestingly, the XAFS results indicate a nearly atomic scale composition fluctuation in this compound, and the consequent {\it ab-initio} calculations based on this chemically inhomogeneous structure succeeds to explain all the unusual electronic and magnetic properties of this compound.

We thank D D Sarma for useful discussions. SJ thanks CSIR, India for fellowship. SR thanks DST-RFBR and DST Fast Track, India for financial support. The work was supported by the Swedish Foundation for International Cooperation in Research and Higher Education within the Institutional Grants programme.

\end{document}